\documentclass[12pt,preprint]{aastex}

\usepackage{psfig}

\def\q{\qquad}
\def\beg{\begin{eqnarray}}
\def\ende{\end{eqnarray}}
\def\lsim{\lower.4ex\hbox{$\;\buildrel <\over{\scriptstyle\sim}\;$}}
\newcommand{\Ha}{\mbox{Ha}}
\newcommand{\Pm}{\mbox{Pm}}
\newcommand{\Rm}{\mbox{Rm}}
\newcommand{\Rey}{\mbox{Re}}
\renewcommand{\vec}[1]{\mbox{\boldmath $#1$}}
 
\def\Om{{\it \Omega}}
\def\R{{R\"udiger}}
\def\A{Alfv\'en}

\shorttitle{Magnetic instability}
\shortauthors{R\"udiger, M. Schultz  et al.}

\begin{document}

\title{Diffusive  MHD instabilities beyond the   Chandrasekhar theorem}
\author{G\"unther R\"udiger, Manfred Schultz}
\affil{Leibniz-Institut f\"ur Astrophysik Potsdam,
         An der Sternwarte 16, D-14482 Potsdam,\\
	 Helmholtz-Zentrum Dresden-Rossendorf, POB 510119, 01314 Dresden, Germany}
\email{gruediger@aip.de, mschultz@aip.de}
\author{Frank Stefani}      
\affil{Helmholtz-Zentrum Dresden-Rossendorf, POB 510119, 01314 Dresden, Germany}
\email{f.stefani@hzdr.de} 
\author{Michael Mond}     
 \affil{Department of Mechanical Engineering,
Ben-Gurion University, Israel}
\email{mond@bgu.ac.il}        

\date{\today}

\begin{abstract}
The magnetohydrodynamic stability of axially unbounded   cylindrical flows is considered which contain a toroidal magnetic background field with the same radial profile as the linear azimuthal  velocity. 
Chandrasekhar (1956) 
has shown for ideal   fluids the stability of this configuration   if the \A\ velocity of the field  equals the velocity of the background flow. It is demonstrated for  magnetized Taylor-Couette flows   at the Rayleigh line, however,  that for finite diffusivity such  flows become unstable against nonaxisymmetric perturbations  where the critical magnetic Reynolds   number of the rotation rate  does not depend on the   magnetic Prandtl number $\Pm$ if $\Pm\ll 1$. 

In order to study this new diffusive  `azimuthal magnetorotational instability',  flows and fields with the same radial profile but with different amplitudes  are 
considered. For $\Pm\ll 1$ the instability domain with the weakest  fields and the  slowest  rotation rates lies {\em below} the 
Chandrasekhar line of equal amplitudes for \A\ velocity and rotation velocity. We find  that  then the  
 lines of marginal instability scale with the Reynolds number and the Hartmann number. The minimum values of the field strength and the rotation rate which are needed for the 
   instability (slightly) grow for more and more flat rotation. Finally,  the corresponding electric current 
 of the background field  becomes so strong that the Tayler instability (which even exists without rotation) also appears in  the bifurcation map  at small  
  Hartmann numbers displacing after all the azimuthal magnetorotational instability. 
\end{abstract}



\section{Motivation}
Plane and parallel hydrodynamic shear flows are only unstable in the inviscid theory against 
infinitesimal perturbations if their span-wise velocity profile has an inflexion point (Rayleigh 1880). 
There is 
no such inflexion for a plane Poiseuille flow with the profile $1-y^2$ between the walls at $y=\pm 1$ so that they are stable for vanishing viscosity. Plane Poiseuille flows with finite viscosity, however, are unstable against infinitesimal disturbances if their Reynolds number $UL/\nu$ exceeds the (high)  value 5772  (Drazin \& Reid 1981). Such flows are destabilized by the finite diffusivity. If the linear instability is considered as a structure-forming process then only the viscosity gives rise to  the structure in this case. This is opposite to those expectations that  any diffusivity should act against the formation of local and also global maxima and minima. 

Magnetohydrodynamic theory provides yet another wide range of phenomena in which equilibria that are stable under zero magnetic-diffusivity may no longer be stable when finite electrical conductivity is considered (Furth et al. 1963, Coppi et al. 1966).
The resulting resistive instabilities  such as the tearing modes drive many observed phenomena in astrophysics, space as well as laboratory plasmas (Connor et al. 2009, Landi \& Bettarini 2012). Examples include solar flares and coronal mass ejection that are related to the tearing mode driven magnetic reconnection, and limited the operational regimes in Tokamaks.

Such disspation-induced instabilities, which are also well known in quite a
number of finite-dimensional mechanical systems, were comprehensively
surveyed by Krechetnikov \& Marsden (2007).

The role of Rayleigh's inflexion point theorem in hydrodynamics is played in magnetohydrodynamics by a theorem of Chandrasekhar (1956) who stated that the solution 
\beg
\vec{U}=\vec{U}_{\rm A}
\label{chandra1}
\ende
of the MHD equations is {\em linearly stable} for ideal and incompressible flows. Here, $\vec U$ is the flow aligned with the magnetic field $\vec{B}=\sqrt{\mu_0\rho}\vec{U}_{\rm A}$ with $\vec{U}_{\rm A}$ the 
\A\ velocity of the field. The fluid mass density is $\rho$ and the vacuum permeability is $\mu_0$. Tataronis \& Mond (1987) studied the stability of the plasma if one works with 
\beg
\vec{U}=\beta \vec{U}_{\rm A}
\label{chandra2}
\ende
with constant $\beta$ and discussed the destabilizing effects of $\beta\neq 1$. In the present paper  we shall present the destabilizing effects of finite diffusivities (viscosity and/or magnetic diffusivity) for a special realization of the Eqs. (\ref{chandra1}) and  (\ref{chandra2}), i.e.  of 
Taylor-Couette flows of electrically conducting fluids  between rotating concentric cylinders. We shall show that the Chandrasekhar theorem does no longer hold if at least  one of the two diffusivities has a finite value. It is no problem to  find for such fluids unstable solutions even for $\beta=1$. Our work is also motivated by a recent result of Kirillov \& Stefani (2013) 
who  used a short-wave approach to derive in the inductionless
limit an analytical expression for the marginal stability curve in terms of the
steepnesses of the angular frequency and the Alfv\'en frequency. The present paper  may also serve  to probe such WKB results
by more elaborate 1D stability investigations. 

Consider  the interaction of the differential rotation in an axially unbounded   Taylor-Couette container and a toroidal magnetic field between the inner and the outer cylinder   which is maintained by  axial electric currents outside and/or inside the inner cylinder. The fluid possesses the microscopic viscosity $\nu$ and the magnetic diffusivity $\eta=1/\mu_0\sigma$ ($\sigma$ the electric conductivity). The general solution of the stationary and axisymmetric equations is 
\beg
U_\phi=R\Om=a_\Om R+\frac{b_\Om}{R},  \q
B_\phi=a_B R+\frac{b_B}{R},
\label{basic}
\ende
where $a_\Om$, $b_\Om$, $a_B$ and $b_B$ are constants which fulfill the condition (\ref{chandra1}) if
\beg
a_\Om = a_B/\sqrt{\mu_0\rho}, \ \ \ \ \ \ \       b_\Om=b_B/\sqrt{\mu_0\rho}. 
\label{chandra3}
\ende
 The most popular realization of the condition (\ref{chandra1}) is   the  rotating pinch where an  axial and uniform-in-radius electric current is 
 subject to rigid-body rotation where both the azimuthal flow and the azimuthal field linearly depend on the radius $R$  (Acheson 1978, Pitts \& Tayler 1985, \R\ \& Schultz 2010).
Another  very special example of the stability problem 
is formed for $a_\Om=a_B=0$ describing the interaction of the rotation law $\Om\propto1/R^2$ (the Rayleigh limit) with the field $B_\phi\propto 1/R$ which is current-free between the cylinders. These profiles fulfill the condition (\ref{chandra2}) but they become unstable against {\em nonaxisymmetric} perturbations with the azimuthal quantum number $m=\pm 1$ for $\beta=1$ if one of the two diffusivities does not vanish. Because of its current-free character we have called it the azimuthal magnetorotational instability (\R\ et al. 2007, Hollerbach et al. 2010)  which even has been realized in the laboratory with liquid alloy GaInSn  as the conducting fluid (Seilmayer et al. 2014).
\section{Equations}
The solution of the equations are governed by the ratios
\begin{equation}
r_{\rm in}=\frac{R_{\rm{in}}}{R_{\rm{out}}}, \; \; \;
\mu_\Om=\frac{\Om_{\rm{out}}}{\Om_{\rm{in}}},  \; \; \;
\mu_B=\frac{B_{\rm{out}}}{B_{\rm{in}}}.
\label{mu}
\end{equation}
$R_{\rm{in}}$ and $R_{\rm{out}}$ are the radii of the inner and the outer
cylinder, $\Omega_{\rm{in}}$ and $\Omega_{\rm{out}}$ are their rotation
rates and $B_{\rm{in}}$ and $B_{\rm{out}}$ are the azimuthal magnetic fields
at the inner and outer cylinders. Conditions (\ref{chandra1}) and (\ref{chandra2}) are both fulfilled for all $\mu_B r_{\rm in}=\mu_\Om$. For our standard model which works with $r_{\rm in}=0.5$ one finds  $\mu_B =2 \mu_\Om$. In this paper  mainly  the  two rotation laws with $\mu_\Om=0.25$ (i.e. with $\mu_B=0.5$) and  $\mu_\Om=0.35$  (i.e. with $\mu_B=0.7$) are used which describe i) the Rayleigh limit of uniform angular momentum and ii) a quasikeplerian rotation law with cylinders rotating as $R^{-3/2}$  (like planets).   
The governing dimensionless equations
\begin{equation}
\Rey (\frac{\partial \vec{U}}{\partial t} + (\vec{U}\cdot \nabla)\vec{U}) =
- \nabla P +  \Delta \vec{U} + 
\Ha^2{\textrm{curl}}\ \vec{B} \times \vec{B}
\label{mom}
\end{equation}
for the momentum and 
\begin{equation}
\Rm( \frac{\partial \vec{B}}{\partial t}- {\textrm{curl}} (\vec{U} \times \vec{B}))=  \Delta\vec{B}
\label{mhd}
\end{equation}
for the induction
are linearized with 
$
{\textrm{div}}\ \vec{U} = {\textrm{div}}\ \vec{B} = 0$  and numerically solved for no-slip boundary conditions and for insulating and/or perfect-conducting cylinders which are unbounded in axial direction.  Those boundary conditions are applied at both $R_{\rm{in}}$ and $R_{\rm{out}}$. 
The dimensionless free  parameters in (\ref{mhd})  are  the Hartmann number ($\Ha$) and the Reynolds number ($\Rey$), given by
\beg
{\rm{Ha}}=\frac{B_{\rm{in}} D}{\sqrt{\mu_0 \rho \nu \eta}},  \ \ \ \ \ \ \  
{\rm{Re}}=\frac{\Om_{\rm{in}} D^2}{\nu},
\label{pm}
\ende
where $D=R_{\rm{out}}-R_{\rm{in}}$ is the unit of length which  is here always $D=R_{\rm out}/2$. With the magnetic Prandtl number ${\rm{Pm}} = {\nu}/{\eta}$  the magnetic Reynolds number of the rotation is $\Rm=\Pm\ \Rey$. The Lundquist number of the magnetic field is ${\rm S}=\sqrt{\Pm} \Ha$. Also the modified magnetic Reynolds number 
\beg
\overline{\Rm}=\sqrt{\Rey \Rm}
\label{Rmq} 
\ende
as a counterpart of the Hartmann number will here play an important role. The reason is that the ratio $\overline{\Rm}/\Ha$ which defines the parameter $\beta$ in (\ref{chandra2}) does not depend on the values of the diffusivities. The code which solves the above equation system is described in detail by  R\"udiger et al. (2014) where also the detailed  formulation of the boundary conditions can be found. The cylinders of the TC container can be assumed as perfect-conducting and/or as insulating. In the present paper 
we mainly but not always  applied vacuum boundary conditions to the magnetic fields. Test calculations have shown that our  basic numerical findings  do not depend on the choice of the bounday conditions. All the minima of the instability curves in the $\Ha$-$\Rey$ plane and the characteristic meeting points of the $\Ha$ axis for resting cylinders exist for both sorts of boundary conditions, mostly the numerical values of the characteristic Reynolds  and  Hartmann numbers  for conducting  cylinders exceed those   for insulating ones.

\section{The  Rayleigh limit}
\begin{figure}[ht]
\psfig{figure=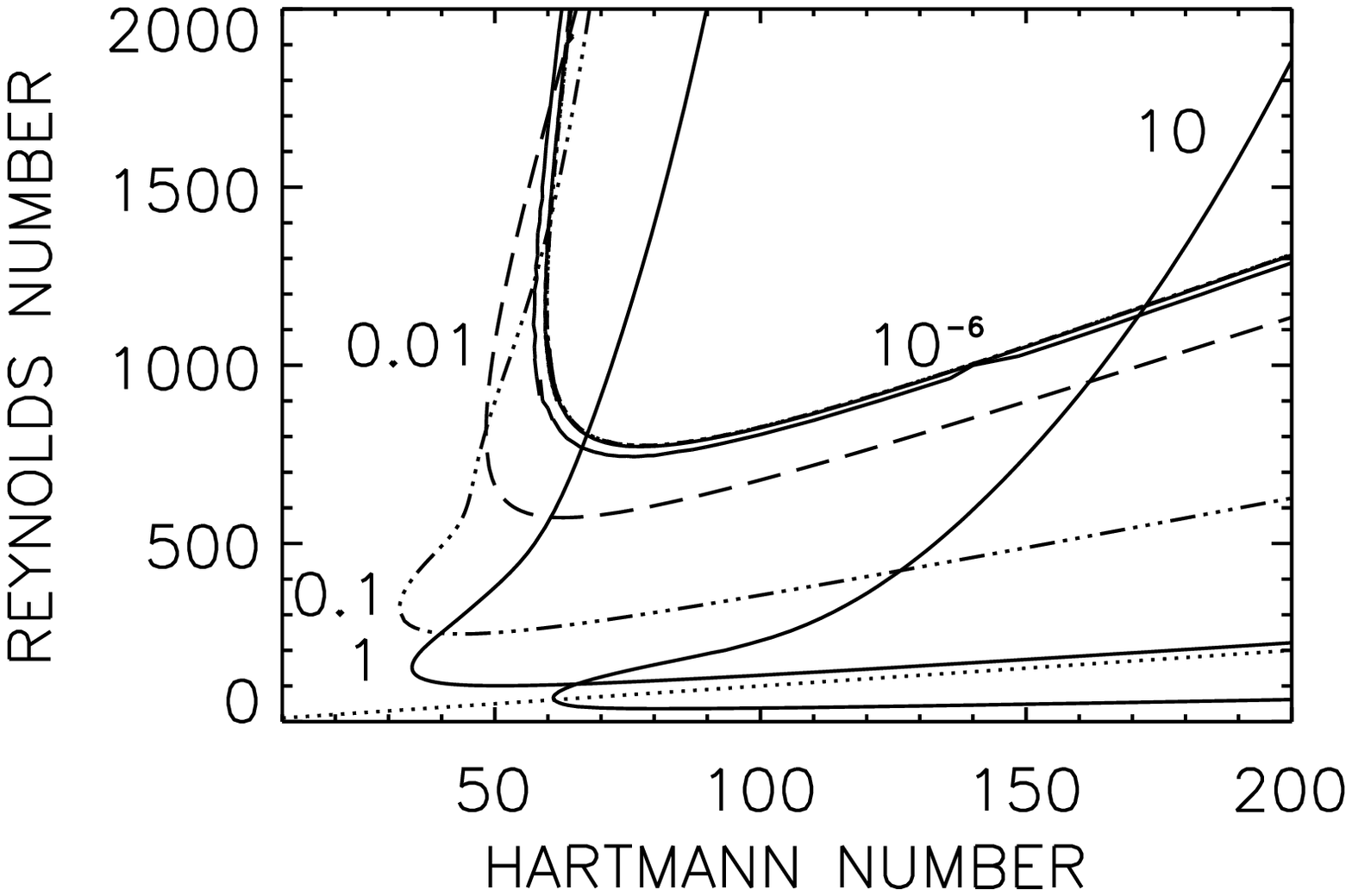,width=8cm}
\psfig{figure=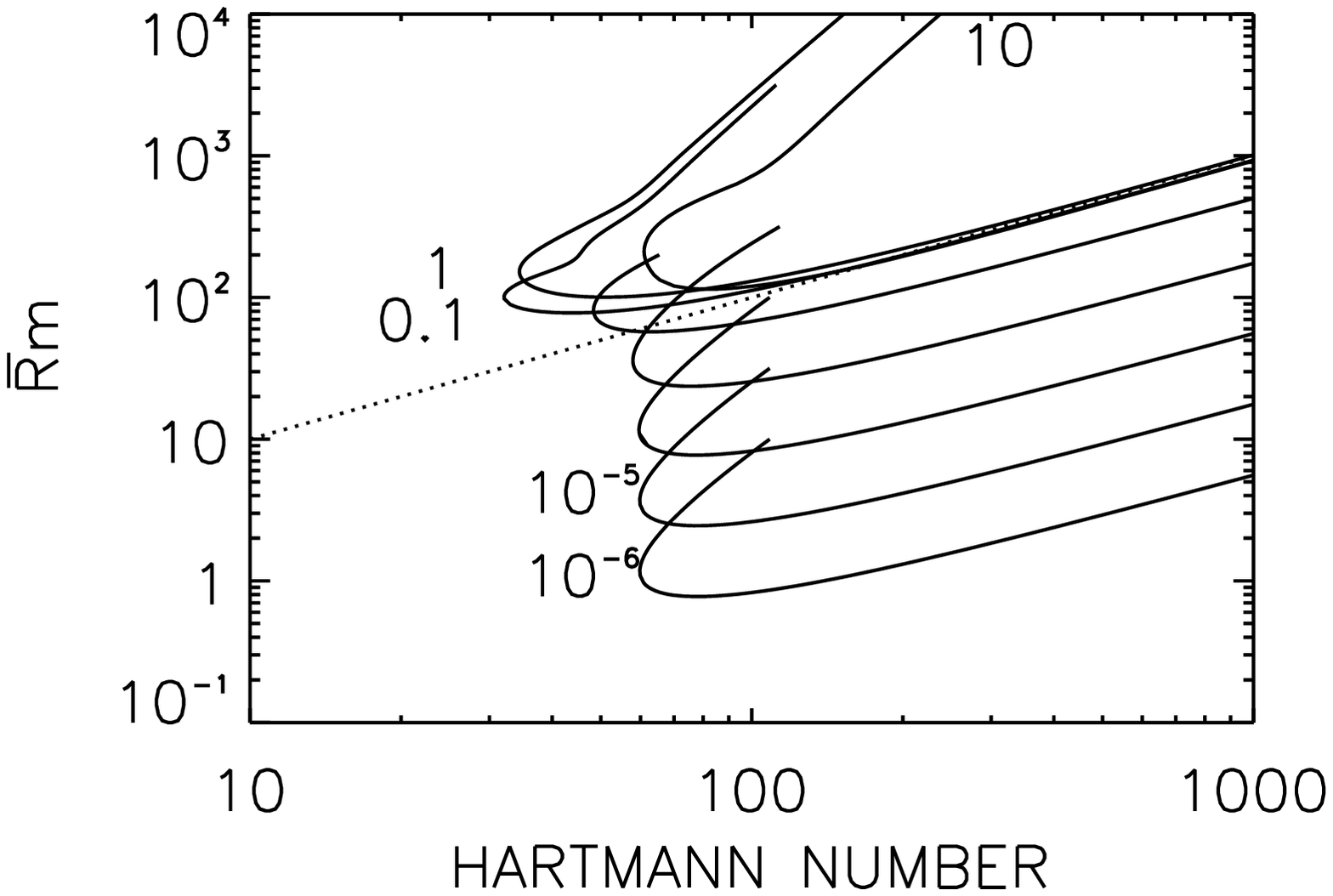,width=8cm}
\caption{\label{fig1} The lines of marginal instability of the nonaxisymmetric modes with $m=\pm 1$ for $\mu_\Om=\mu_B/2=0.25$. The curves are marked with their magnetic Prandtl numbers. The curves are plotted in the $\Rey-\Ha$ plane (top) and in the  $\overline{\Rm}-\Ha$ plane (bottom). The values along the dotted line  fulfill the stability condition (\ref{chandra1}). Note that for small enough $\Pm$ the minimum Hartmann numbers for instability do hardly depend on the magnetic Prandtl number. }
\end{figure}

Figure \ref{fig1} shows the lines of marginal stability for the  rotation law with $U_\phi\propto B_\phi\propto 1/R$ (i.e. $\mu_\Om=0.25$,  $\mu_B=0.5$). For a given supercritical Hartmann number the instability always  exists between   a minimum Reynolds number and a maximum Reynolds number.  The lower branch of the instability cone defines the critical rotation rate which is necessary to provide the needed energy for the pattern maintenance. The upper branch limits the instability domain by suppressing the nonaxisymmetric instability by too strong shear. If, however, the applied azimuthal magnetic field contains too strong axial electric-currents then the lower branch degenerates to a vertical line which for $ \Rey=0$ crosses the $ \Ha$-axis at the characteristic Hartmann number $\Ha_{\rm Tay}$ (which does not depend on the magnetic Prandtl number
(R\"udiger \& Schultz 2010).

For very  small $\Pm$  the curves converge and form a common minimum Reynolds number at a certain  critical Hartmann number   (Fig. \ref{fig1}, top).  The value of the minimum Reynolds number decreases for growing magnetic Prandtl number but the smallest critical Hartmann number is reached for $\Pm$ of order unity. For very small $\Pm$ the minimum of the instability cone scales with $\Rey$, here with a value of order 800  while the typical Hartmann number is ten times less.

By use of the modified Reynolds number $\overline{\Rm}$ one gets another  picture.
  The dotted line in Fig. \ref{fig1}  (bottom) is defined by   $\overline{\Rm}=\Ha$ representing  the location of all values fulfilling (\ref{chandra1}). It is {\em always} crossed by all the bifurcation lines for finite $\Pm$. For these cases the flow is unstable even under the condition (\ref{chandra1}). 

\begin{figure*}[ht]
\hbox{
\psfig{figure=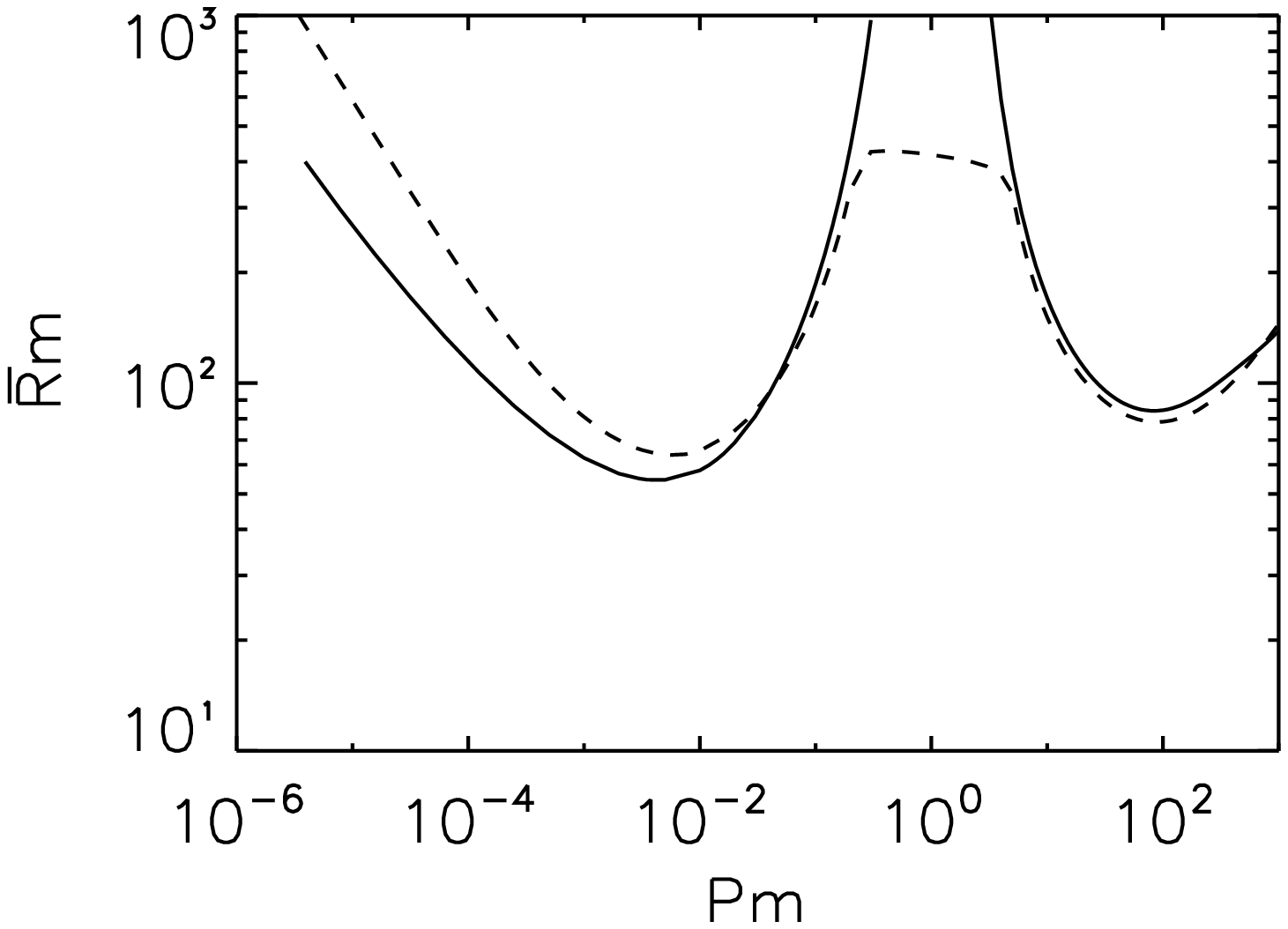,width=5.5cm}
\psfig{figure=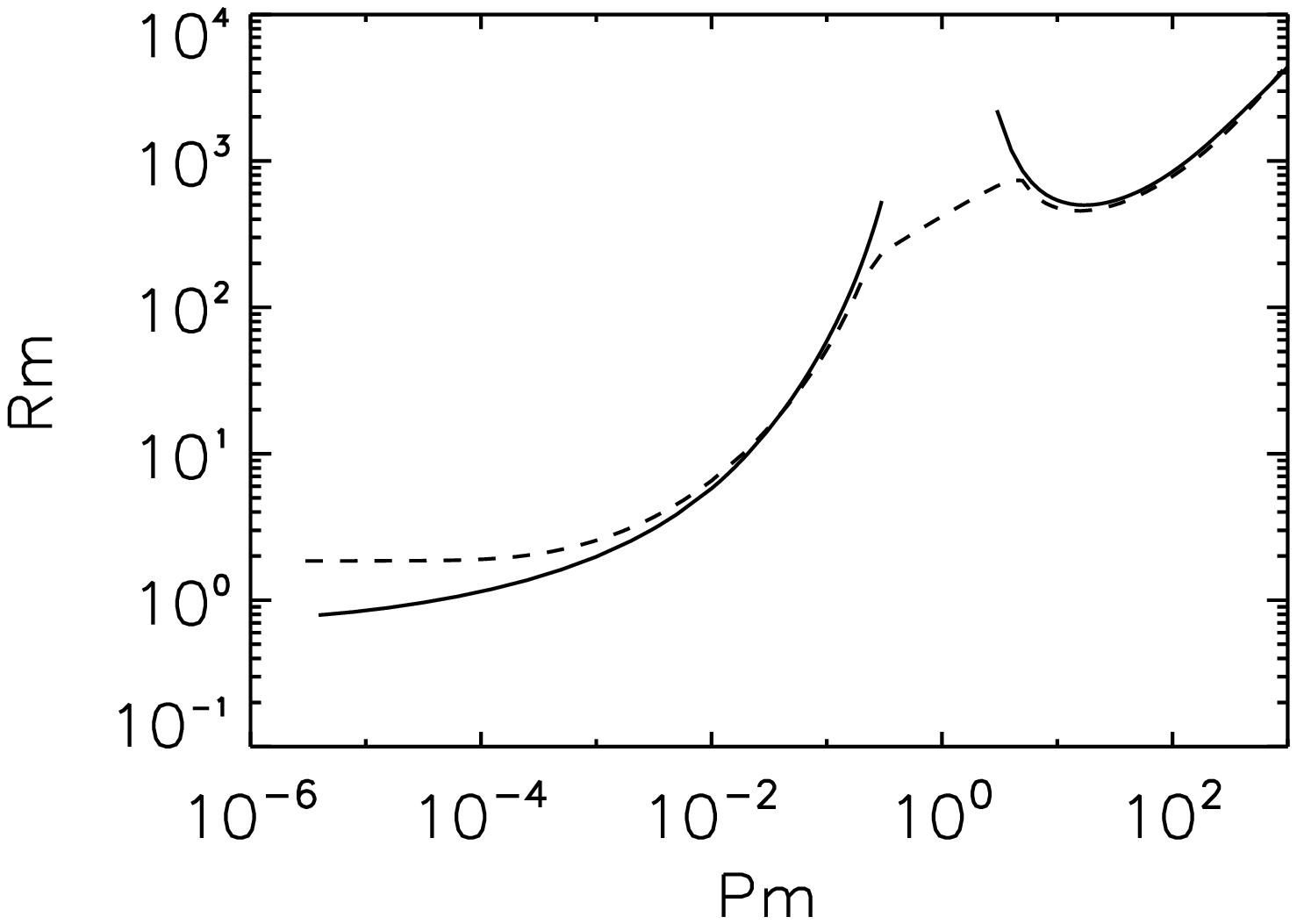,width=5.5cm}
\psfig{figure=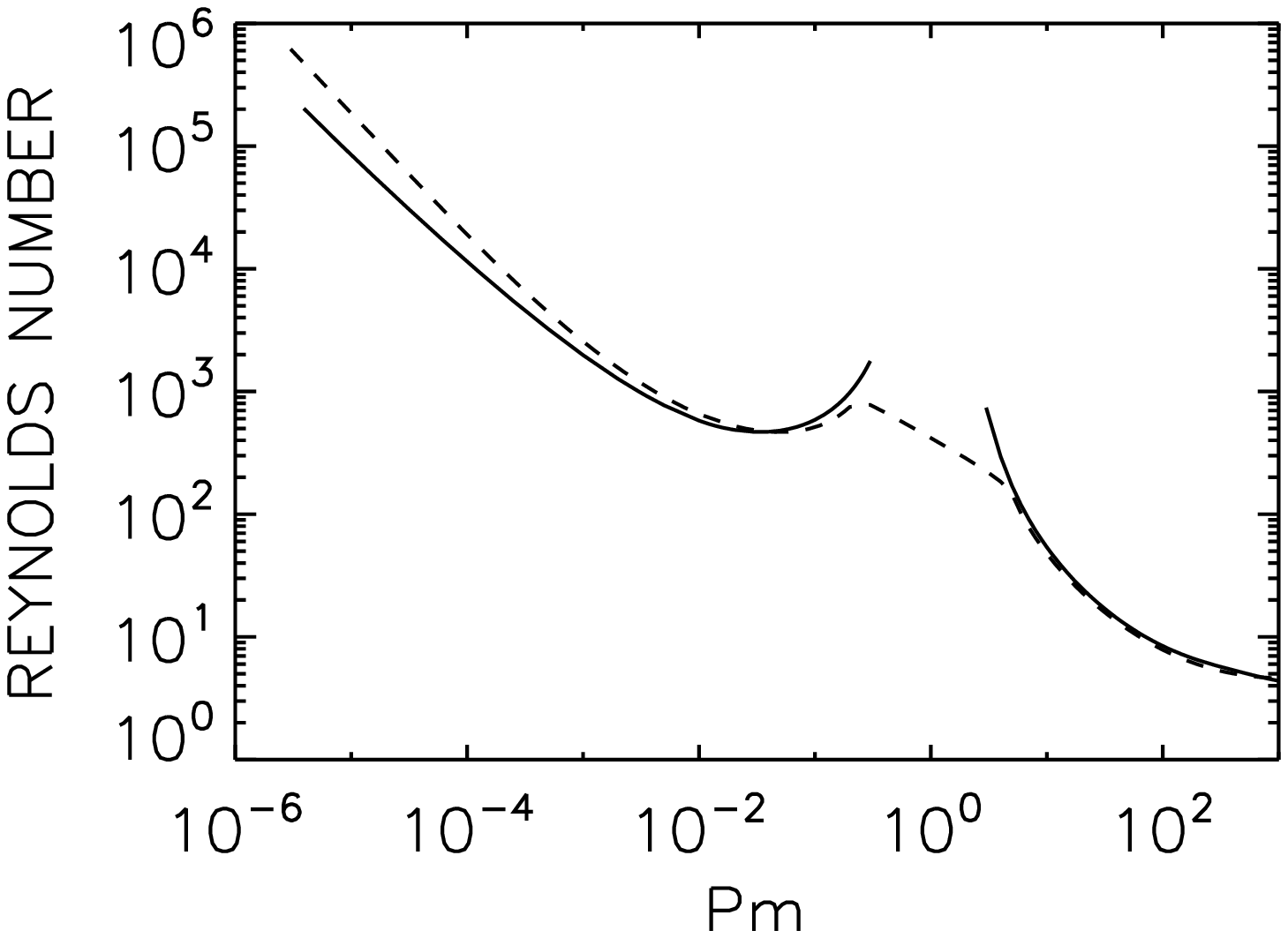,width=5.5cm}}
\caption{\label{fig2} Left: The location of the eigenvalues after (\ref{chandra1})  vs. the magnetic Prandtl number $\Pm$ for conducting-boundary conditions  (solid) and insulating boundaries (dashed). 
Middle: For very small $\Pm$ it scales with $\Rm$. Right:  For very  large $\Pm$ it scales with $\Rey$.}
\end{figure*}

The numerical values of the crossing points are plotted in  Fig. \ref{fig2}. The solid (dashed)  line corresponds to models with perfect-conducting (insulating) cylinders. Both cases lead to very similar results. In the  $\overline{\Rm}-\Ha$ plane one finds minimal Hartmann numbers for small and for large $\Pm$. For $\Pm=1$ the curves have a local maximum which reflects the phenomenon that the main part of the cones for $\Pm>1$ lies above the dotted line in Fig. \ref{fig1} (bottom) while it lies below the dotted line for $\Pm<1$. For both models the Hartmann number for $\Pm=1$ reaches large values but remains finite. 

On the other hand, the Hartmann numbers and -- what is the same -- the Reynolds numbers $\overline{\Rm}$ of the crossing points run with $\Pm^{-1/2}$ for $\Pm\to 0$ which means that the magnetic Reynolds number $\Rm$ of the rotation and the Lundquist number $\rm S$ of the magnetic field remain finite. Figure \ref{fig2} demonstrates that the solutions for $\Pm\to 0$ possess values of $\Rm\simeq 0.8$ for perfect-conducting cylinders and  $\Rm\simeq 2$  for insulating cylinders. Similar results could be formulated  for $\Pm\to \infty$ but for the Reynolds number $\Rey$ (Fig. \ref{fig2}, right). We have thus shown for the rotation law of the Rayleigh limit that  even for the case that only one of the two diffusivities  is nonvanishing unstable modes exist along the line  defined by  the stability condition  (\ref{chandra1}).
\section {Quasikeplerian rotation}
\begin{figure}[ht]
\psfig{figure=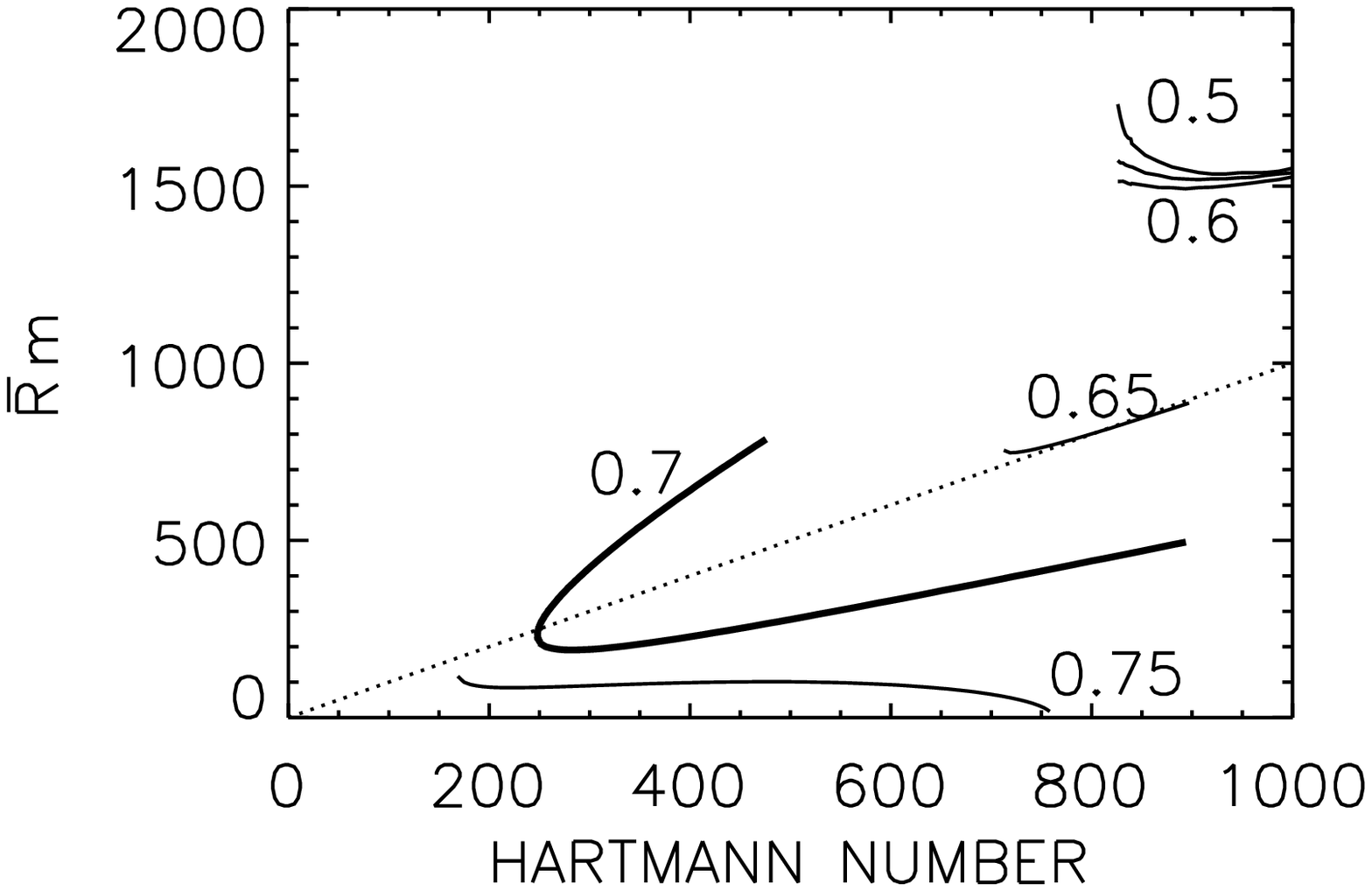,width=8cm}
\caption{\label{fig3}   The instability maps of the nonaxisymmetric modes with $m=\pm 1$ for $\Pm=0.001$. The dotted line represents the condition (\ref{chandra1}). The bold curve values are fulfilling (\ref{chandra2}). The curves are labeled with their  $\mu_B$-values. $\mu_\Om=0.35$.}
\end{figure}

\begin{figure}[ht]
\psfig{figure=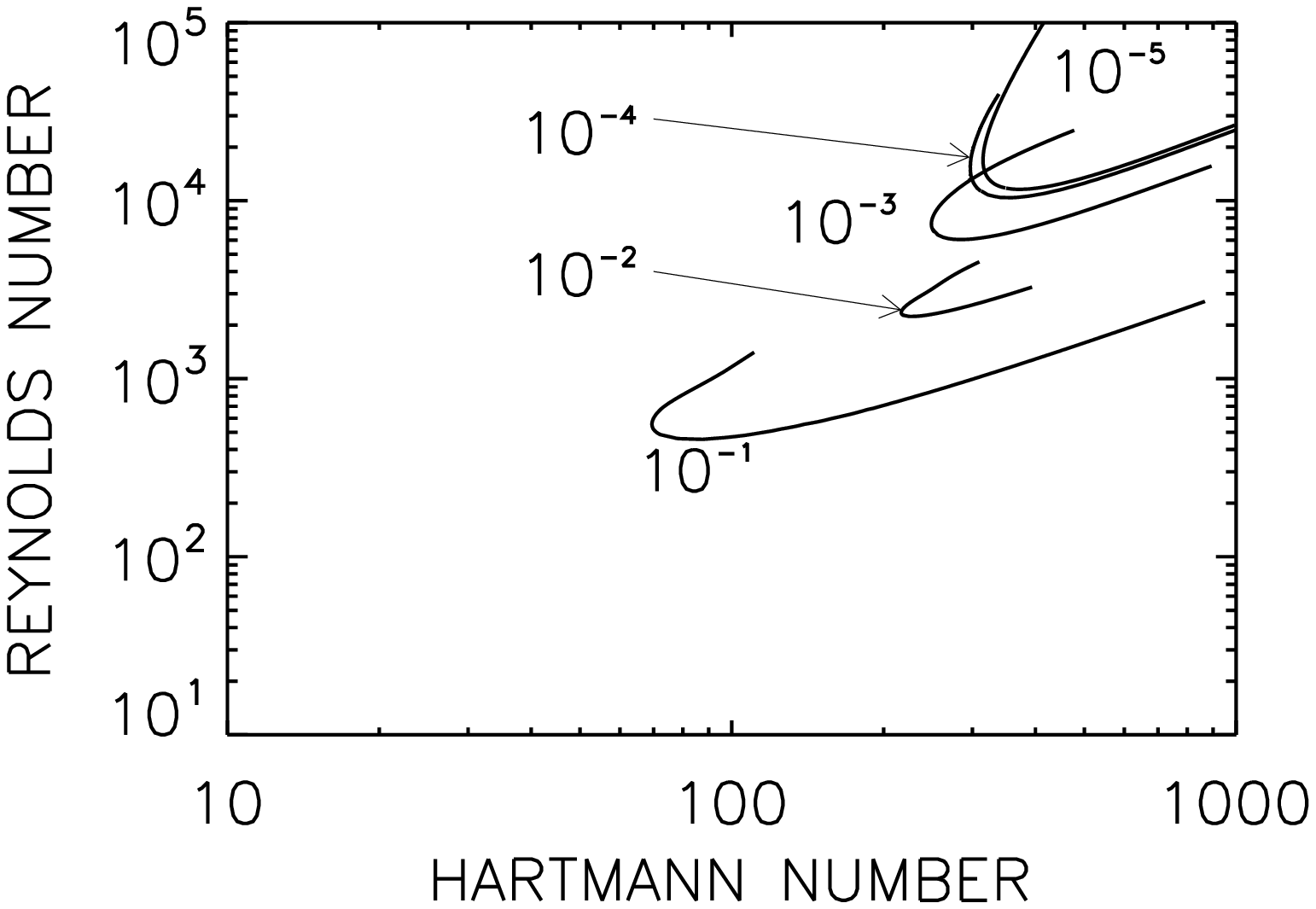,width=8cm}
\psfig{figure=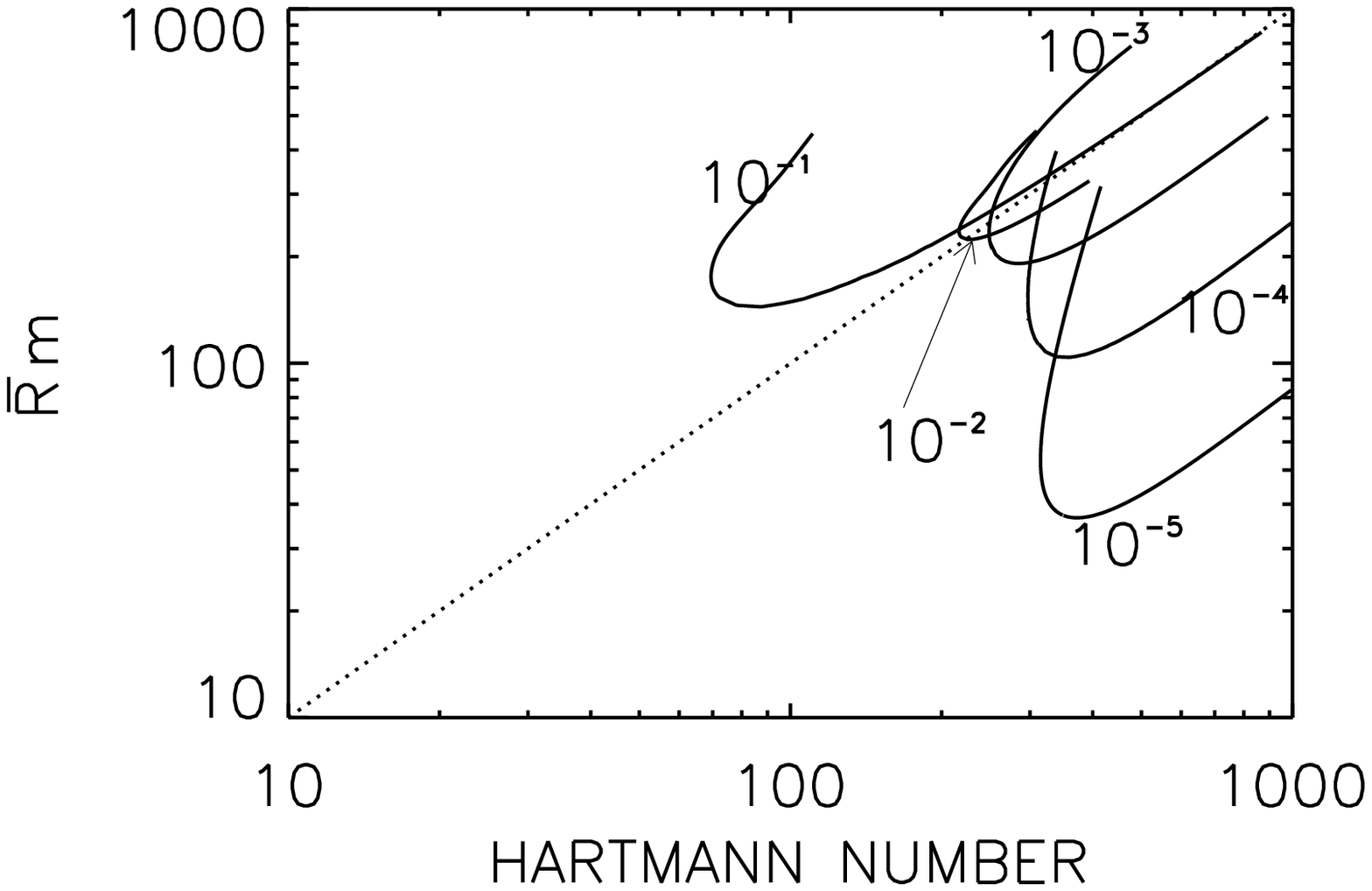,width=8cm}
\caption{\label{fig4} The same as in Fig. \ref{fig1}  but for the quasikeplerian rotation law with $\mu_\Om=\mu_B/2=0.35$. For small enough $\Pm$ the minimum Hartmann numbers for instability do hardly depend on the magnetic Prandtl number. The curves are plotted in the $\Rey-\Ha$ plane (top) and in the  $\overline{\Rm}-\Ha$ plane (bottom). Vacuum boundary conditions. For conducting boundary condition the the minimum of the curves for $\Pm\to 0$ scales with  $\Rey\simeq 9\cdot 10^4$ and  $\Ha\simeq 800$.}
\end{figure}

The toroidal fields whose radial profiles  differ from  the above case with  $B_\phi\propto R^{-1}$  can only be maintained by use of  electric currents in axial direction between the cylinders.  The immediate consequence is the appearance of a kink-type Tayler instability for  resting cylinders ($\Rey=0$). 
Figure \ref{fig3} shows the bifurcation maps of the instability of the nonaxisymmetric modes with $m=\pm 1$ for  fixed magnetic Prandtl number for various radial profiles of the azimuthal magnetic field but for fixed (quasikeplerian) rotation law. 
The dashed lines represent the Rayleigh profile ($B_\phi\propto R^{-1}$)  and the Kepler 
profile ($B_\phi\propto R^{-3/2}$). All the curves except the first one ($\mu_B=0.5$) meet the horizontal axis at finite 
values  $\Ha_{\rm Tay}$ which do not depend on the value of  $\Pm$ (see \R\ et al. 2013).
  Our model yields $\Ha_{\rm Tay}=2565$ for $\mu_B=0.7$ and  $\Ha_{\rm Tay}=760$ for $\mu_B=0.75$ (shown). Differential rotation included the minimum Hartmann number for instability are much less  than these values ($\Ha_{\rm min}=250$ for $\mu_B=0.7$,  see Fig. \ref{fig3}). Also the minimum Hartmann numbers necessary for excitation only slightly depend on the magnetic Prandtl numbers  if the latter  are  small enough  (Figs. \ref{fig1} and \ref{fig4}). These minimum Hartmann numbers for quasikeplerian rotation {\em are  significantly higher} than those for the rotation at the Rayleigh limit.

The instability cones  in the $\Rey-{\rm Ha}$-plane for the quasikeplerian radial profiles 
with $\mu_\Om=\mu_B/2=0.35$ are given in Fig. \ref{fig4} (top) for various magnetic Prandtl 
numbers in correspondence to   Fig. \ref{fig1} (top).  Again for small magnetic Prandtl numbers  
the instability cones do not depend on $\Pm$, they thus {\em again} scale with the Reynolds number $\Rey$ and the Hartmann number $\Ha$. The minimum
 Hartmann number and minimum Reynolds number  for the Kepler profiles exceed the 
 corresponding values for the Rayleigh-limit profiles by almost one order of magnitude. 
 It is known, however, that the quasikeplerian rotation ($\mu_\Om=0.35$) together with the 
 current-free magnetic field ($\mu_B=0.5$) for small $\Pm$ scales with the magnetic Reynolds 
 number $\Rm$. Obviously, the  electric current in the axial direction extends the range of the shear beyond 
 the Rayleigh limit to higher values where the instability curves scale with $\Rey$.

The same instability domains  in the $\overline{\rm Rm}-{\rm Ha}$-plane  are given in 
Fig. \ref{fig4} (bottom) such as in  Fig. \ref{fig1} (bottom)  for the Rayleigh-limit  profile. 
Again for small $\rm Pm$ the minimum Hartmann number and minimum Reynolds number  
for the Kepler profiles exceed the corresponding values for the Rayleigh-limit profiles. 
Except for these differences the basic schemes of the instability maps of Fig. \ref{fig4} are  the 
same  as in Fig. \ref{fig1}. For decreasing $\Pm$ the cones migrate downward 
crossing in one point the Chandrasekhar line (\ref{chandra1}).   
Each of these crossing points  represents a  marginal unstable solution  which fulfills the 
stability condition (\ref{chandra1}) for ideal fluids.
\begin{figure}[ht]
\psfig{figure=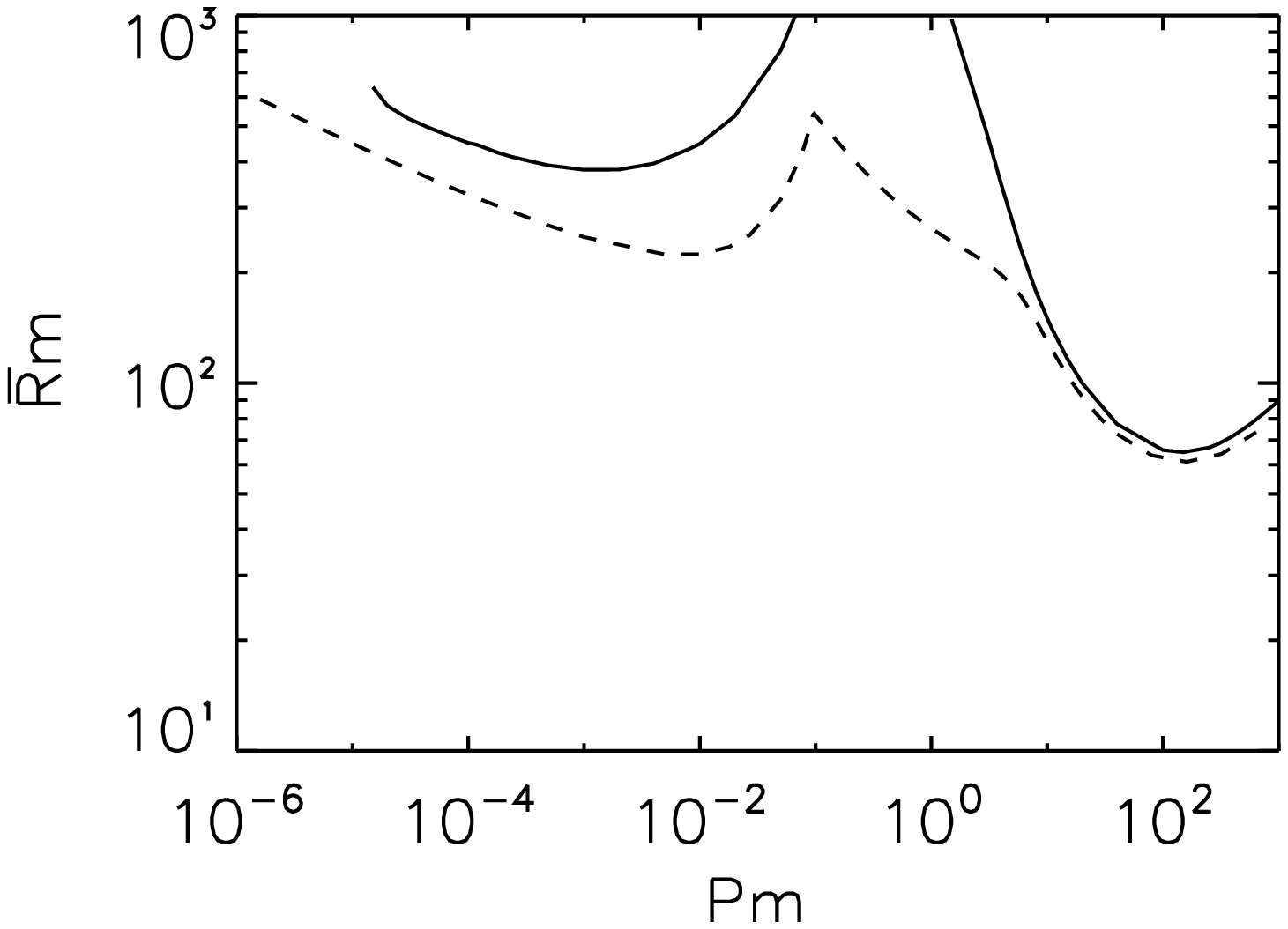,width=8cm}
\psfig{figure=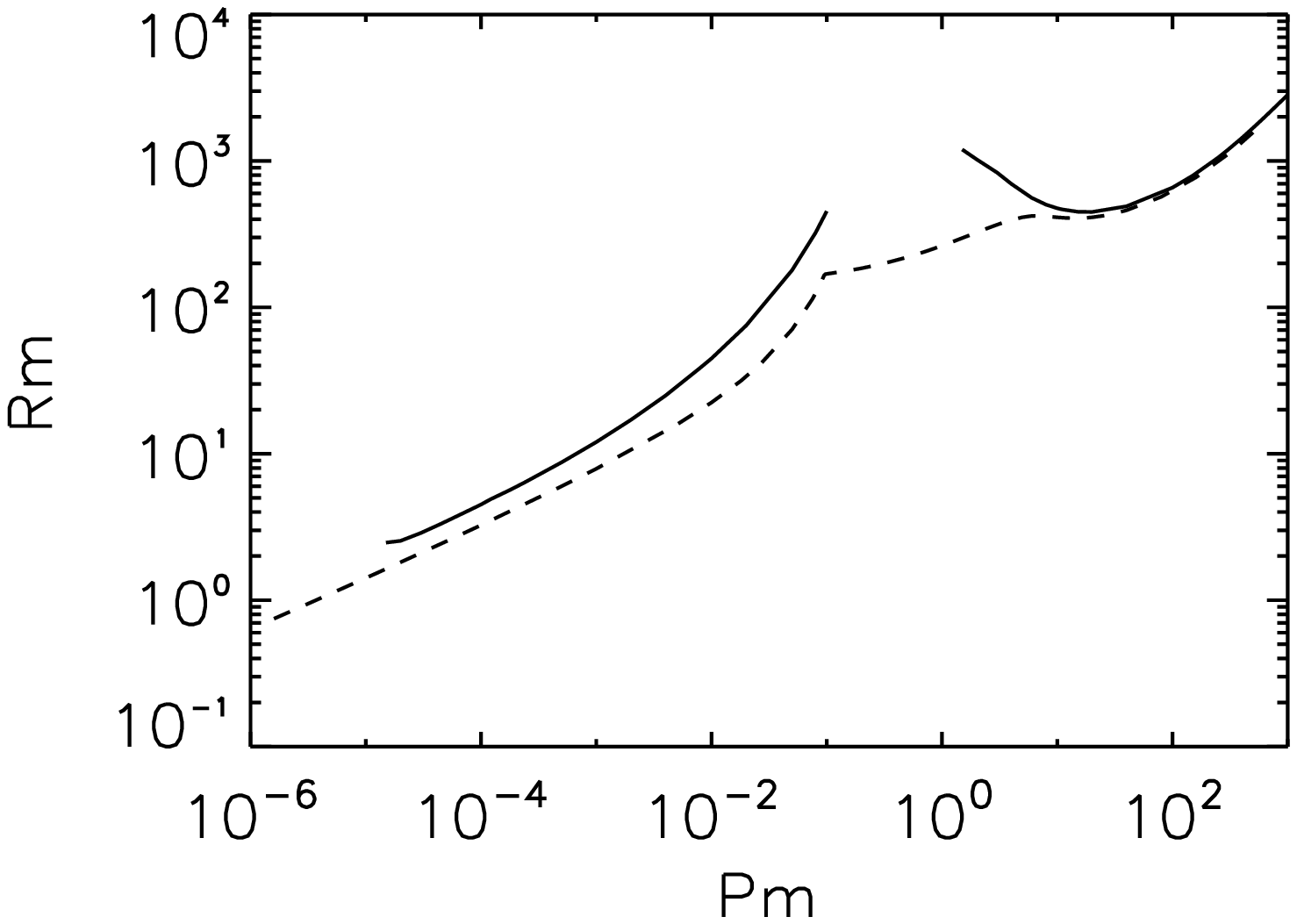,width=8cm}
\caption{\label{fig5} The same as in Fig. \ref{fig2} but for  the quasikeplerian rotation law with $\mu_\Om=\mu_B/2=0.35$. Conducting-boundary conditions  (solid), insulating boundaries (dashed).}
\end{figure}

As  in Fig. \ref{fig2} in Fig. \ref{fig5} the coordinates of the crossing points are plotted in dependence on the magnetic Prandtl number for both sets of boundary conditions. In opposition to the situation at the Rayleigh line for $\Pm\to 0$ neither $\overline{\Rm}$ nor $\Rm$ show finite values. 
One finds  $\Rm\propto \Pm^{1/3}$, hence for small    $\Pm$ 
\beg
\frac{\Om_{\rm{in}} D^2}{\sqrt[3]{\nu\eta^2}}\simeq 100.
\label{kep}
\ende
There is a characteristic difference, therefore, of the behavior of the crossing points for $\Pm\to 0$.
While the magnetic Reynolds number for the flow at the Rayleigh limit remains finite, it slowly 
vanishes for  quasikeplerian rotation. The behavior of the critical rotation rate 
for $\nu\to 0$ is thus opposite:  it remains finite for the Rayleigh flow but it becomes 
infinitely small for the Kepler flow. Note , however, by comparison of the Figs. \ref{fig2} 
and \ref{fig5}  that for $\Pm$ of order $10^{-5}$ or $10^{-6}$ (the values for fluid metals 
like sodium or gallium) the magnetic Reynolds number and the Hartmann number of the 
crossing points in both cases are almost equal.  This is also approximately true for  higher magnetic Prandtl number up to  $\Pm\lsim 10^{-1}$.
\section{Beyond the Kepler limit}
\begin{figure}[ht]
\psfig{figure=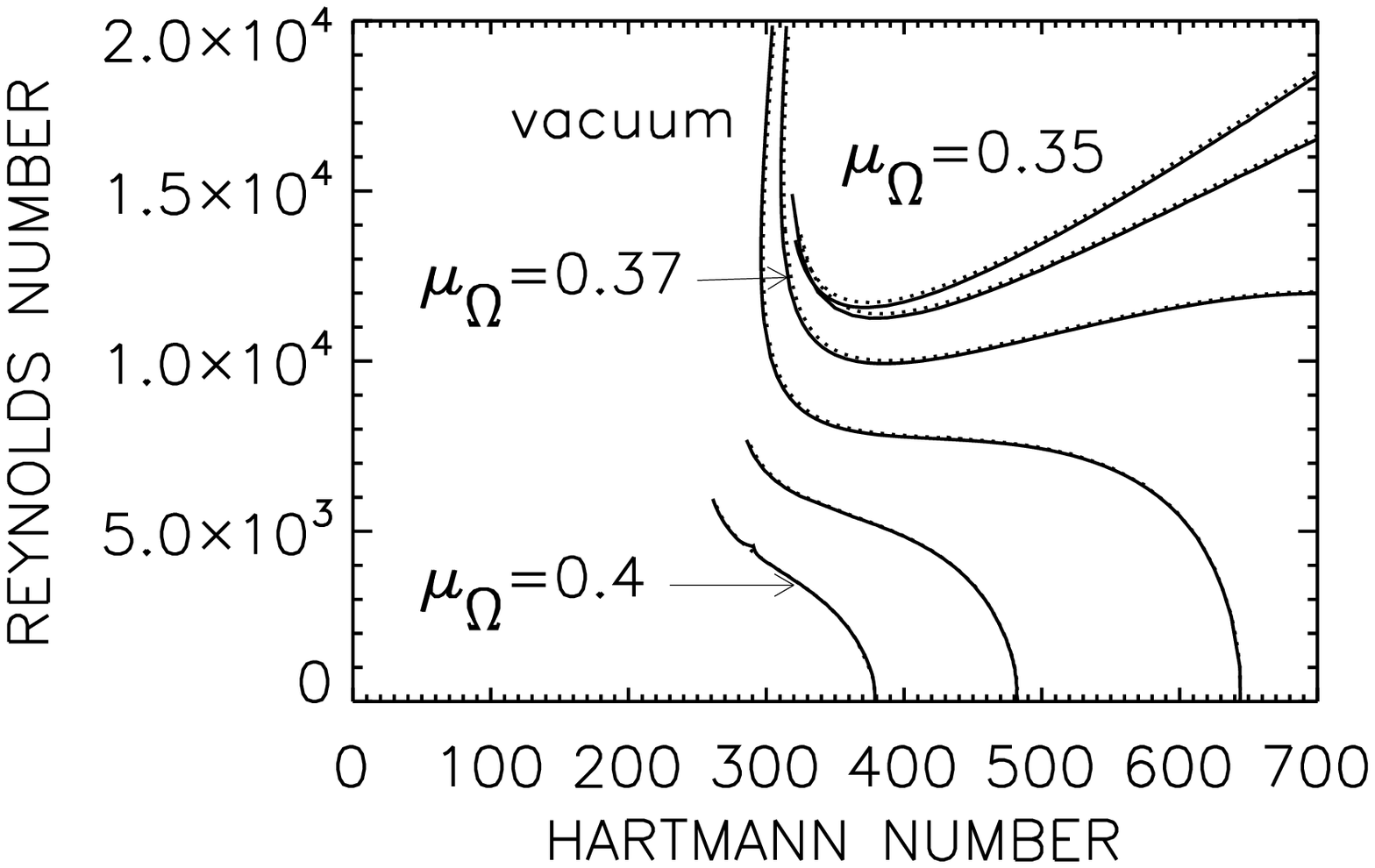,width=8cm}
\caption{\label{fig6} Critical Reynolds numbers   for the modes $m=\pm 1$ for the  rotation laws with $\mu_\Om=0.35...0.40$. $\mu_B=2 \mu_\Om$, insulating boundaries, $\Pm=10^{-5}$, $\Pm=10^{-6}$ (dotted).}
\end{figure}

With the Fig. \ref{fig6} we can show that also for the  rotation law with $\mu_\Om=0.37$ the instability scales with the Reynolds number for $ \Pm\to 0$. This rotation law is  flatter than the Kepler law (`subkeplerian').  For  $ \Pm\to 0$ the location of the characteristic instability cones in the $\rm Re-Ha$-plane looses any  dependence on the magnetic Prandtl number $\Pm$. The same is true  for the previous examples with 
 $\mu_\Om=0.25 $  and  $ \mu_\Om=0.35 $. Hence,  if both  azimuthal flow and azimuthal field  fulfill Eq. (\ref{chandra2}) --  for $\Pm\to 0$ the location of the instability domains in the $\rm Re-Ha$-plane does  {\em not} depend on the magnetic Prandtl number -- or with other words-- the instability scales with Hartmann and Reynolds number.  One  can easily  show that all magnetohydrodynamic equations which possess solutions for $\Pm\to 0$ (which is not identical to $\nu=0) $ basically  scale with $\rm Ha$ and $\Rey$.

The instability domain in Fig. \ref{fig6} forms a cone which is opened to both the large values of $\Rey$ and $\Ha$ or -- with other words -- the line of marginal instability exhibits  a minimum and two branches with positive slope. Almost always the rotation can thus be too slow or too fast and also the magnetic field can be too weak or  too strong for the instability. However, 
all models with $\mu_B>0.5$ contain an axial electric current within the fluid which becomes Tayler-unstable for no or for slow  rotation (Tayler 1973). The line of marginal instability for these modes  thus always meets the horizontal coordinate 
axis for $\Rey=0$ at a critical Hartmann number $\Ha_{\rm Tay}$. The latter becomes smaller for increasing electric current.  As Fig. \ref{fig6} shows, already for $\mu_\Om\simeq 0.38$ 
the $\Ha_{\rm Tay}$ becomes so close to the minimum  that it    disappears. The slope of the lower branch of the instability cone becomes negative so that the requirement of a 
minimum critical rotation rate no longer  exists. It is obvious that the higher amplitude of the electric current changes the character of the instability towards the character of the (rotation-influenced) Tayler instability. 

\begin{figure}[ht]
\psfig{figure=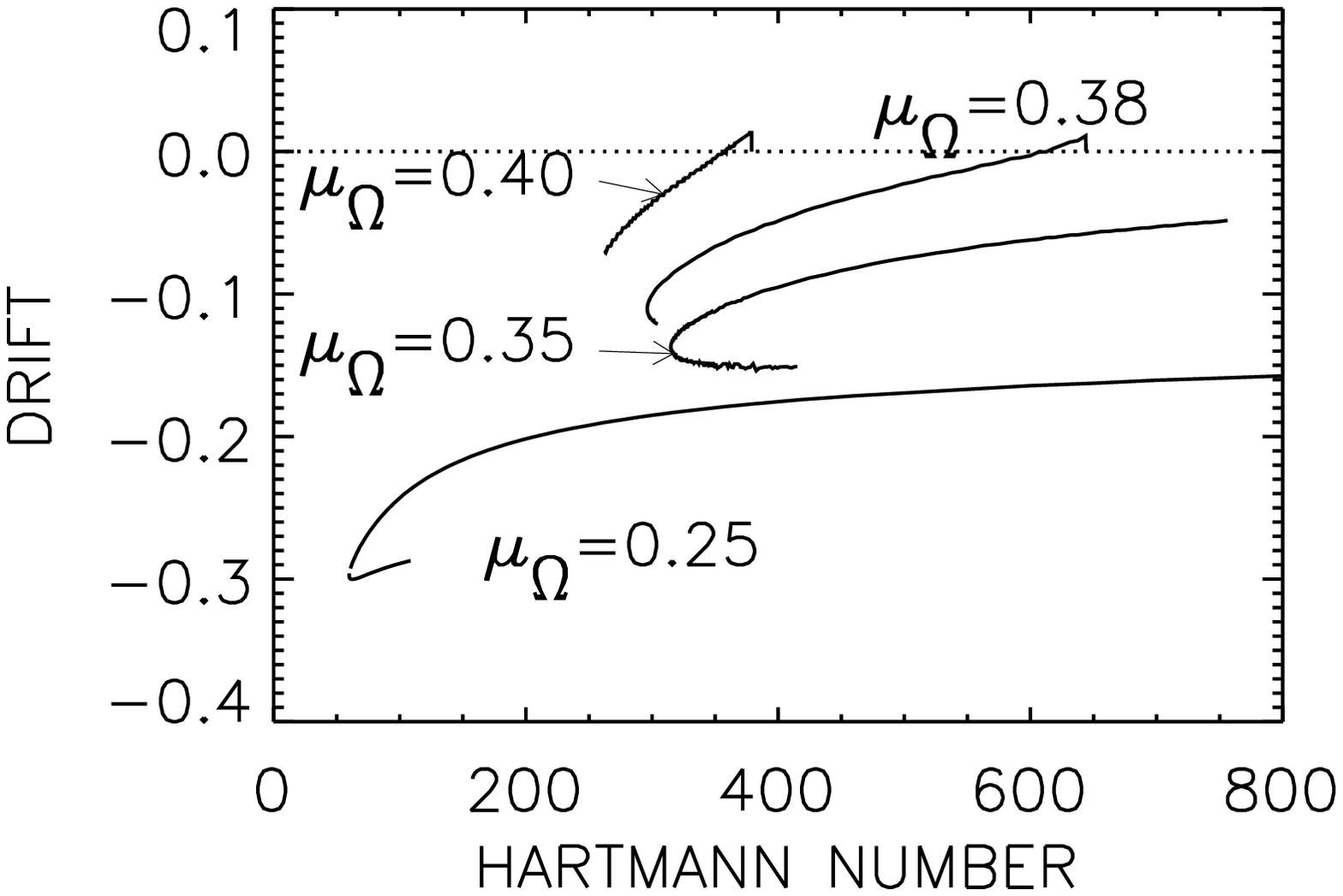,width=8cm}
\caption{\label{fig8} Drift rates (normalized with the inner rotation rate) for the modes $m=\pm 1$ for the  rotation laws with $\mu_\Om=0.25...0.40$. The dotted line denotes the solutions which rest in the laboratory system. $\mu_B=2 \mu_\Om$, insulating boundaries, $\Pm=10^{-5}$ and $\Pm=10^{-6}$.}
\end{figure}

The drift rates are also reflecting  the change of the instability characteristics for growing electric-current amplitudes. While for small $\Pm$ the magnetic pattern  for the Rayleigh rotation ($\mu_\Om=0.25$) and the Kepler rotation ($\mu_\Om=0.35$) migrates in the direction of the global rotation it rests in the laboratory system for the subkeplerian  rotation laws with $\mu_\Om> 0.35$  (see Fig. \ref{fig8}). Both extrema are known: the pattern of AMRI at the Rayleigh line tends to  rotate with the outer cylinder while the Tayler instability without rotation basically rests in the laboratory system (Seilmayer et al. 2012). 

Nevertheless, for small $\Pm$ the lines of marginal instability in the $\Ha-\Rey$~plane do not depend on   the magnetic Prandtl number also for the branches in Fig. \ref{fig6} with negative slope which cross the axis $\Rey=0$. It was already known that the values of $\Ha_{\rm Tay}$ (i.e. the critical Hartmann number for the Tayler instability without rotation) do not depend on $\Pm$ (see also R\"udiger et al. 2013). Obviously, the same is true  for electric-currents subject to  differential rotation for small $\Pm$. All the excitation conditions which we derived   for models with $U_\phi\propto B_\phi$ scale with $\Rey$ and $\Ha$ for  $\Pm\to 0$.

\section{Conclusions}
  Galaxies are the only cosmical objects whose internal flows and fields can simultaneously be observed. Their magnetic fields of several  tens of $\mu$gauss with densities of 10$^{-24}$ g/cm$^3$ correspond to an \A-velocity of  about 100 km/s which numerically complies with the condition (\ref{chandra1}) as the linear velocity of the galactic rotaton is of the same order. Flows and fields which strictly   fulfill the condition (\ref{chandra1})  are stable in ideal fluids.
Our numerical study of azimuthal fields in a differentially rotating Taylor-Couette container shows, however,  that this magnetohydrodynamic configuration can become unstable against nonaxisymmetric perturbations in fluids with finite diffusivities.  This dissipation-induced instability is a perfect
illustration of Montgomery's  (1993) verdict '...that
for fluid equations of the Navier-Stokes type
the ideal limit with zero dissipation coefficients
has essentially nothing to do with the case of
small but finite dissipation coefficient'.
In particular,  at the Rayleigh limit with the rotation law  $\Om\propto R^{-2}$ even  
the existence of one of the two diffusivities enables the 
 the  instability to occur. One finds  for $\Pm\to 0$ the excitation condition 
 \beg
\frac{\Om_{\rm{in}} D^2}{\eta}\simeq 0.8
\label{kep1}
\ende
(the numerical value valids for conducting cylinders) while for $\Pm\to \infty$ a similar condition holds with $\eta$ replaced by $\nu$. This one-diffusivity phenomenon is restricted to the Rayleigh limit as  for flatter rotation laws the condition for $\Pm\to 0$ reads different (see Eq.  (\ref{kep})). 
Note that after (\ref{chandra1}) the rotation at the  Rayleigh limit corresponds  to a magnetic profile $B_\phi\propto 1/R$ which is the only one in the considered   cylindric geometry which is maintained by electric currents which completely flow outside the fluid. 

If the condition (\ref{chandra1}) is replaced by the condition (\ref{chandra2}) with $\beta$ independent of $R$ then for $\beta\neq 1$ and for
 fixed $\Pm$ one finds that the eigenvalues  defined by  (\ref{chandra1}) belong to an infinite and smooth  line of marginal instability with two branches with different but positive slope in the $\Ha-\Rey$~plane. Each of the curves possess a  minimum  Hartmann number and a minimum Reynolds number (which lie close together).  For $\Pm\ll 1$ the  the curves in the $\Ha-\Rey$~plane do not depend on the magnetic Prandtl number (the curves `scale' with $\Rey$ and $\Ha$). 

As already described, the scaling with   $\Rey$ and $\Ha$ for small $\Pm$ is already known for AMRI at the Rayleigh limit but it is new that {\em all} configurations which fulfill (\ref{chandra2}) show the same behaviour (see Fig. \ref{fig6}).  Also the instability lines for much flatter rotation laws, or what is here the same, for more flat magnetic profiles (which  need  axial electric current also within the fluid for their maintenance) do not depend on the value of $\Pm$ if $\Pm\ll 1$.
 
  It is clear that for $\mu_B>0.5$ and sufficiently strong field amplitude  the bifurcation lines meets the $\Ha$~axis as the axial currents are unstable against nonaxisymmetric  perturbations for $\Rey=0$. We know that the corresponding Hartmann number $\Ha_{\rm Tay}$ does not depend on the magnetic Prandtl number of the fluid. Rigid rotation suppresses the TI while differential rotation with $\mu_\Om<1$ acts supporting. Hence, for slow rotation the instability lines above the $\Ha$~axis always turn to the left. Also  these positions do not depend on $\Pm$ --  but only if  $\Pm\ll 1$.  
  
  The critical Hartmann number $\Ha_{\rm Tay}$ is decreasing for increasing  $\mu_B$. For $\mu_B=1$ it is $\Ha_{\rm Tay}=109$ (150) for vacuum (conducting) boundary conditions.  These values are so small that the formation of an AMRI minimum is no longer possible. In this case the AMRI disappears and the resulting bifurcation line takes the form  of the bifurcation line of the Tayler instability under the influence of the given  differential rotation (see Fig. \ref{fig6}). 
  
  One can also recognize the transition from AMRI to the Tayler instability  for increasing $\mu_B=2 \mu_\Om$ by means of the drift frequency of the magnetic nonaxisymmetric pattern which develops from corotation with the outer cylinder (for AMRI) to resting in the laboratory system (see Fig. \ref{fig8}).
  
  Our  basic result, however, is that for fields and flows with the same radial profile  all the described bifurcation lines in the $\Ha-\Rey$~plane do not depend on $\Pm$ if  $\Pm\ll 1$. As also the current-free field ($\mu_B=0.5$) together with the rotation law at the Rayleigh limit 
  ($\mu_\Om=0.25$) belongs to this class it is not surprising that the characteristic minimum of the curve of marginal instability scales with $\Ha$  and  $\Rey$  even for the smallest magnetic Prandtl numbers  which has been   experimentally confirmed  (Seilmayer 2014). This is no longer  true, however,  if flow and field cannot be expressed by condition (\ref{chandra2}) as for example for   current-free fields  ($\mu_B=0.5$) under the influence of  Kepler rotation  ($\mu_\Om=0.35$) where the scaling of the minima can only be expressed by the Lundquist number of the field and the magnetic Reynolds number of the rotation which for small $\Pm $
   leads to experimentally
  unrealistic high values of the magnetic field and the rotation rate.



\end{document}